\documentclass[twocolumn,prl,a4paper,aps,showpacs,amsmath,amssymb]{revtex4}
\usepackage{graphicx}
\usepackage{dcolumn}
\usepackage{bm}

\begin{document}

\title{Giant Isotope Effect in the Incoherent Tunneling Specific Heat of the Molecular Nanomagnet Fe$_{8}$}

\author{M. Evangelisti$^{1,2}$, F. Luis$^{3}$, F. L. Mettes$^{1}$, R. Sessoli$^{4}$, and L. J. de Jongh$^{1}$}

\affiliation{$^{1}$ Kamerlingh Onnes Laboratory, Leiden University, 2300 RA Leiden, The Netherlands\\ $^{2}$
National Research Center on ``nanoStructures and bioSystems at Surfaces'' (S$^{3}$), INFM-CNR, 41100 Modena,
Italy\\ $^{3}$ Instituto de Ciencia de Materiales de Arag\'{o}n, CSIC-Universidad de Zaragoza, 50009 Zaragoza, Spain\\
$^{4}$ Dipartimento di Chimica, Universit\`{a} di Firenze, 50144 Firenze, Italy}
\date{\today}

\begin{abstract}
Time-dependent specific heat experiments on the molecular nanomagnet Fe$_{8}$ and the isotopic enriched
analogue $^{57}$Fe$_{8}$ are presented. The inclusion of the $^{57}$Fe nuclear spins leads to a huge
enhancement of the specific heat below 1~K, ascribed to a strong increase in the spin-lattice relaxation rate
$\Gamma$ arising from incoherent, nuclear-spin-mediated magnetic quantum tunneling (MQT) in the
ground-doublet. Since $\Gamma$ is found comparable to the expected tunneling rate, the MQT process has to be
inelastic. A model for the coupling of the tunneling spins to the lattice is presented. Under transverse
field, a crossover from nuclear-spin-mediated to phonon-induced tunneling is observed.
\end{abstract}

\maketitle

\newpage

Single-molecule magnets are fascinating nano-size superparamagnetic particles which at low temperatures may
flip their magnetic moments by magnetic quantum tunneling (MQT) through the anisotropy
barrier~\cite{Gatteschi03}. Observation of quantum tunneling in these molecules illustrates the complexity of
the interaction of such magnetic qubits with their ``environment'' (neighboring particles, nuclear spins,
phonons). Indeed, the tunnel splitting $\Delta$ of the magnetic ground-state is many orders of magnitude
smaller than the energy bias $\xi$ from, e.g., dipolar interactions between molecules, rendering MQT
impossible at first sight. It is by now well established, both theoretically~\cite{Prokof'ev98,Fernandez03}
and experimentally~\cite{Sangregorio98,Wernsdorfer00,Mettes01}, that {\em incoherent} MQT is yet possible
through the presence of rapidly fluctuating nuclear spins. The resulting dynamical hyperfine bias may, at any
time, bring a fraction of the molecular spins into resonance, thus opening an energy window $E_{w}\gg\Delta$
for incoherent tunneling.

It should be emphasized that in the Prokof'ev/Stamp (PS) model~\cite{Prokof'ev98}, relaxation of the magnetic
moment of the molecular cluster (hereafter, cluster spin) is to the nuclear spin bath. A coupling to the
lattice is not considered, the argument being~\cite{Prokof'ev96} that only at long times relaxation by
phonons will become more efficient than the nuclear-spin-mediated magnetic relaxation. However,
time-dependent specific heat experiments~\cite{Mettes01,Evangelisti04} neatly show that (for high enough MQT
rate) both the electronic and nuclear spin systems can remain in thermal equilibrium with the lattice even
deep into the quantum regime, where the only fluctuations possible are those arising from MQT events. This
strongly suggests that, whereas the nuclear spins are needed to relax the (otherwise blocked) electron spins
through MQT, at the same time the MQT mechanism apparently enables relaxation of both nuclear and electron
spins to the lattice. Interestingly, in magnetic insulating compounds such as these, relaxation of nuclear
spins to the lattice has to occur via the electron spin-phonon channel, direct nuclear relaxation to the
lattice being extremely slow at low temperatures. In other words, by enabling the cluster spins to tunnel,
the nuclei themselves can relax to the lattice!

In this Letter we present definite proof for this unusual scenario. By comparing the low-$T$ specific heat
arising from incoherent tunneling in the ground-doublet for Fe$_{8}$ and its $^{57}$Fe-enriched counterpart
$^{57}$Fe$_{8}$, we show that the inclusion of the $^{57}$Fe nuclear moments in the (otherwise identical)
molecules leads to an enormous enhancement of the specific heat below 1~K, which can only result from a
strong increase in the spin-lattice relaxation rate $\Gamma$. Below 1~K, this rate is found to be several
orders of magnitude larger than predicted for conventional spin-lattice relaxation~\cite{Politi95}. To
explain our results an extension of the PS model is presented that includes inelastic tunneling events, in
which a spin-flip is accompanied by the creation or annihilation of low-energy phonons.

Low-temperature $(0.1~{\rm K}<T<7~{\rm K})$ specific heat $C$ measurements were performed in a home-made
calorimeter~\cite{Mettes01} using the thermal relaxation method. By varying the thermal resistance of the
link between calorimeter and cold-sink, the characteristic timescale $\tau_{e}$ of the experiment can be
varied. In this way measurements of the time-dependent $C$ can be exploited to probe the long-time
($0.1-10^{2}$~s) magnetic relaxation~\cite{Mettes01}. At higher temperatures ($T>2$~K), $C$ was measured
using a commercial calorimeter. The enriched sample, denoted by $^{57}$Fe$_{8}$, was prepared as described in
Ref.~\cite{Wernsdorfer00}. Except for the presence of the $^{57}$Fe nuclear moments in $^{57}$Fe$_{8}$
(enriched to 95\% in $^{57}$Fe), no difference in the magnetic structure between Fe$_{8}$ and $^{57}$Fe$_{8}$
is to be expected.

\begin{figure}[t!]
\centering{\includegraphics[angle=0,width=8.6cm]{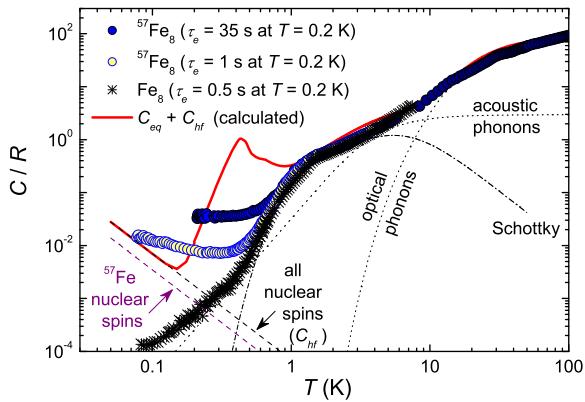}} \caption{(color online) Zero-field specific heats
of non-oriented samples of standard Fe$_{8}$ and of $^{57}$Fe$_{8}$ as a function of temperature. For
$^{57}$Fe$_{8}$, data for $\tau_{e}\approx 35$~s ($\bullet$) and 1~s ($\circ$) are given, whereas for
Fe$_{8}$, $\tau_{e}\approx 0.5$~s ($\ast$). Drawn curves are explained in the text.}
\end{figure}

We first discuss the zero-field specific heat data. In Fig.~1, we compare the specific heat $C/R$ of a
non-oriented powder $^{57}$Fe$_{8}$ sample, as measured with two different thermal links, with our previous
data~\cite{Mettes01} on non-oriented Fe$_{8}$. The specific heat contains contributions from the phonon modes
of the crystal lattice as well as magnetic contributions from electronic and nuclear spins. The lattice
specific heat can be described by the sum of a Debye curve, which describes the contribution of acoustic
phonon modes, plus an Einstein oscillator term that probably arises from intramolecular vibrational modes
(optical phonons). The overall fit yields for both samples $\theta_{D}\simeq 19$~K and $\theta_{E}\simeq
38$~K for the Debye and Einstein temperatures, respectively~\cite{lattice}, (plotted in Fig.~1 as dotted
curves).

Above $1~$K, the magnetic contribution $C_{m}$ to the specific heat of both compounds is independent of
$\tau_{e}$. In this range, the equilibrium $C_{m}$ mainly arises from transitions between the energy levels
of the molecular spin $S=10$, split by the uniaxial anisotropy \cite{Mettes01,Luis98b}. The associated
multilevel Schottky anomaly $C_{0}$ is shown as the dash/dotted curve in Fig.~1. Below $1$~K, the remaining
entropy arises almost completely from the ground-state doublet $\pm 10$, which is split by intercluster
magnetic dipolar couplings and by hyperfine interactions between the electronic and nuclear spins. Dipolar
interactions can induce, under equilibrium conditions, a long-range ordered magnetic phase. In Fig.~1, we
plot the equilibrium specific heat $C_{eq}$ of Fe$_{8}$, obtained from Monte-Carlo
calculations~\cite{Fernandez00}, which predicts the occurrence of such a phase transition at $T_{C}\simeq
0.43$~K. However, the equilibration of the relative populations of the two lowest levels, either by thermal
activation or by tunneling, is a very slow process~\cite{Luis98b}, thus $C_{m}$ measured at finite $\tau_{e}$
does not reach its equilibrium value. Indeed, as shown in Fig.~1 (see also~\cite{Mettes01}), the specific
heat deviates from equilibrium below a blocking temperature $T_{B}\simeq 1.2$~K, indicating that the
electronic spin-lattice relaxation rate $\Gamma$ becomes smaller than $\tau_{e}^{-1}$.

In this non-equilibrium regime we observe a spectacular isotope effect. For the standard Fe$_{8}$, $C_{m}\sim
C_{0}$ below $T\sim 1$~K, decreasing exponentially to nearly zero (below $\sim 0.4$~K the measured $C$
basically equals the Debye term). By contrast, $C_{m}$ of $^{57}$Fe$_{8}$ is about $100$ times larger and it
increases with $\tau_{e}$. We mention here that $\tau_{e}$ varies weakly with $T$ and is determined for each
data point. The $\tau_{e}$ values mentioned in Figs.~1 and 2 are those estimated at $T=0.2$~K for each
particular $T$-sweep. The strong isotope effect is a direct evidence that $\Gamma$, i.e. the thermal contact
between the lattice and electronic spin systems, is enhanced as a result of the introduction of $^{57}$Fe
nuclear spins. Although the rate is not yet sufficient to ensure complete thermal equilibrium within the
experimental time constants available, a sizable part of the entropy of the ground-doublet is now removed.

It should be added that the $^{57}$Fe nuclear spins $I=1/2$ will also contribute to the zero-field $C/R$. In
thermal equilibrium, this contribution amounts to $C_{hf}(^{57}{\rm Fe})/R=A^{2}s^{2}I(I+1)/3(k_{B}T)^{2}$,
where $A$ is the hyperfine coupling and $s=5/2$ is the Fe$^{3+}$ electronic spin. Taking $A/k_{B}=1.65$~mK
estimated by Stamp and Tupitsyn~\cite{Stamp04}, we obtain the $T^{-2}$-term indicated by the lower dashed
curve in Fig.~1. The higher dashed curve in Fig.~1 gives the equilibrium nuclear $C_{hf}$ calculated by
adding the contributions of nuclear spins at the $120$ protons, the $18$ $^{14}$N nuclei and the $8$
$^{79,81}$Br nuclei present in the Fe$_{8}$ molecule. Note that at $T\approx 0.2$~K, $C_{hf}$ is one to two
orders of magnitude smaller than the measured $C$ for $^{57}$Fe$_{8}$. Further, see for standard Fe$_{8}$
that $C$ becomes smaller than $C_{hf}$ at $T\approx 0.1$~K, suggesting that also nuclear spins are off
equilibrium.

We next discuss experiments performed under applied magnetic field. Figure~2 shows results for
$^{57}$Fe$_{8}$ measured at $T\simeq 0.22$~K, compared to previous data for Fe$_{8}$ obtained for nearly the
same $T$ and $\tau_{e}$~\cite{Mettes01}. At such high-$B$ and low-$T$, $C_{m}$ of a randomly oriented sample
is dominated by the contribution of those crystals whose anisotropy axes lie nearly perpendicular to the
field. To substantiate this statement, we calculated that, at $T=0.2$~K and $B=1.5$~T, those crystals making
an angle smaller than $87.5^{\circ}$ with the field contribute less than one percent to the electronic
equilibrium $C$. These experiments give thus information on the way the spin-lattice relaxation is modified
by $B_{\perp}$.

In agreement with the zero-field-behavior, the $C_{m}$ curves of the two isotopic derivatives are very
different for $B<1.5$~T (Fig.~2). For higher fields, however, they are seen to merge within the experimental
uncertainties. The observed dependence on the applied field can be well explained in terms of a tunable
quantum tunneling rate. The perpendicular field introduces off-diagonal terms in the spin Hamiltonian, which
increase the tunnel splitting $\Delta$ by many orders of magnitude. When $\Delta$ becomes larger than
$E_{w}$, the tunneling rate becomes no longer determined by the hyperfine interactions but rather by
$\Delta$. Then $\Gamma$ should become nearly the same for both compounds, as observed. Besides
phonon-assisted tunneling, the increase of $\Delta$ also leads to strong enhancement of the relaxation rate
associated with `conventional' direct processes of emission and absorption of phonons. This is shown in the
inset of Fig.~2, where we plot $\Gamma(B_{\perp}$) calculated~\cite{note} using the anisotropy parameters
from Ref.~\cite{Wernsdorfer99}. For $B_{\perp}\gtrsim 2.2$~T, $\Gamma$ becomes of the same order of the
experimental $\tau_{e}^{-1}$. This explains why conventional theory for spin-lattice relaxation accounts
quantitatively for the transition to equilibrium observed at large transverse fields~\cite{Mettes01},
although it completely fails to account for the zero field data.

\begin{figure}[t!]
\centering{\includegraphics[angle=0,width=8.6cm]{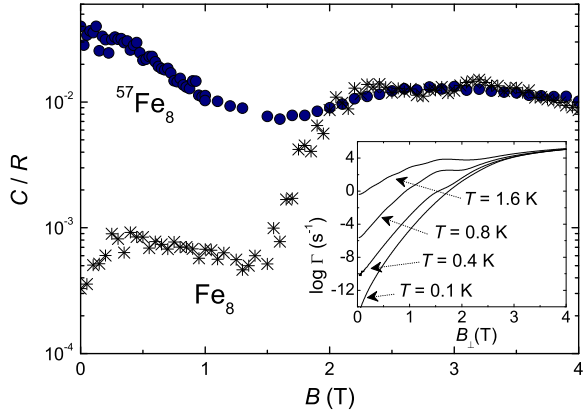}}\caption{(color online) Field-dependence of the
specific heat of $^{57}$Fe$_{8}$ measured at $T=0.22$~K, (for $\tau_{e}\approx 35$~s), together with previous
data~\cite{Mettes01} obtained for Fe$_{8}$ at $0.24$~K. Inset: Calculated $\Gamma(B_{\perp})$ associated with
`conventional' direct processes (see text) for several $T$.}
\end{figure}

To obtain quantitative information on the rate at which the spins approach thermal equilibrium at low $T$ in
zero field, we need to assume a particular expression for the time-dependence of $C_{m}$. This is complicated
by the fact that, at low $T$ and especially below the ordering temperature $T_{C}$, relaxation to equilibrium
becomes a collective process in which each spin-flip modifies the dipolar biases acting on the other
spins~\cite{Prokof'ev98}. This problem was theoretically studied by Fern\'andez~\cite{Fernandez02}, who
calculated numerically the time-dependent $C_{m}$ of a lattice of {\em interacting} Ising spins flipping by
quantum tunneling. Within this model, $C_{m}\simeq C_{0}+(\tilde{c}~T_{C}^{2}/T)(\Gamma/v)$ where $v\equiv
dT/dt$ is the temperature sweeping-rate and $\tilde{c}$ is a constant that depends on the symmetry and
lattice parameters. In our experiments, $v\simeq \Delta T/\tau_{e}$, where $\Delta T\simeq 0.05~T$ is the
$T$-change of the calorimeter in each data point. It is therefore possible to determine $\Gamma$, up to a
constant factor, from $C_{m}$. Moreover, above $T_{C}$ we can also fit $C_{m}$ assuming a simple exponential
decay, $C_{m}(t)=C_{eq}+(C_{0}-C_{eq})\exp(-t\Gamma)$ putting $t=\tau_{e}$, where $C_{eq}$ is the calculated
equilibrium $C$ of the electron spins (see Refs.~\cite{Luis98b,Mettes01} for details). As shown in Fig.~3,
the rates obtained by these two methods overlap, giving $\tilde{c}=0.05~k_{B}$. Deducing $\Gamma$ in this
manner is however restricted to $T$ regions for which the measured $C_{m}$ is sufficiently large compared to
$C_{0}$ and to the nuclear contributions $C_{hf}$. Unfortunately, this is not the case for standard Fe$_{8}$
below $1$~K.

\begin{figure}[t!]
\centering{\includegraphics[angle=0,width=8.6cm]{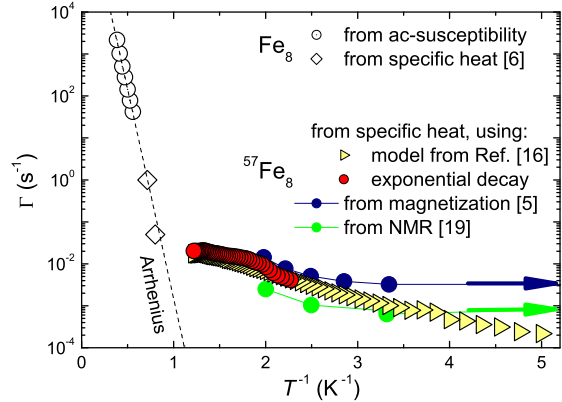}}\caption{(color online) Spin-lattice relaxation
rates of standard Fe$_{8}$ and $^{57}$Fe$_{8}$ obtained by different experimental techniques, as labelled.
Dashed curve is the Arrhenius fit to the high-$T$ data. Arrows indicate low-$T$ limits deduced from
magnetization~\cite{Wernsdorfer00} and NMR~\cite{Baek05} data.}
\end{figure}

Figure~3 shows $\Gamma(T)$ estimated as above together with data obtained at higher temperatures from
Ref.~\cite{Mettes01}. Above 1~K, $\Gamma$ follows an Arrhenius law that is approximately the same for both
Fe$_{8}$ derivatives and corresponds to a thermally activated relaxation over a barrier $U\approx 22.5$~K. By
contrast for $T<1$~K, the decrease of $\Gamma$ of $^{57}$Fe$_{8}$ abruptly slows down and thus $\Gamma$
becomes orders of magnitude faster than predicted for activated behavior. Although we cannot extract $\Gamma$
directly from the low-$T$ $C$ of standard Fe$_{8}$, we can still estimate an upper bound for it. Using the
experimental $\tau_{e}(T)$ of standard Fe$_{8}$ and assuming $\Gamma(T)$ of Fe$_{8}$ to be directly
proportional to $\Gamma(T)$ of $^{57}$Fe$_{8}$, we calculate $C_{m}(T)$ with either the model of
Ref.~\cite{Fernandez02} or exponential decay (see above). The so-obtained $C_{m}(T)$ reproduces well the
experimental $C_{m}(T)$ of standard Fe$_{8}$ if $\Gamma(T)$ is taken a factor of 3 smaller than that of
$^{57}$Fe$_{8}$. The same factor was obtained from time-dependent magnetization
experiments~\cite{Wernsdorfer00}, which provided as well a similar low-$T$ limit of $\Gamma$, also plotted in
Fig.~3. These data have been successfully interpreted in terms of the PS model, i.e. nuclear spin-mediated
tunneling events at a tunneling rate $1/\tau_{t}$, followed by square-root relaxation through redistribution
of dipolar fields throughout the sample in combination with additional spin-flips. In the PS model the
tunneling rate is given by $1/\tau_{t}\approx\Delta_{t}^{2}/E_{w}$. With $\Delta_{t}\approx 5\times
10^{-8}$~K and tunneling window $E_{w}\approx 0.03$~K, as found for $^{57}$Fe$_{8}$~\cite{Wernsdorfer00}, one
obtains $1/\tau_{t}\approx 1\times 10^{-3}~{\rm s}^{-1}$. The square root relaxation rate is basically given
by \cite{Prokof'ev98b}: $1/\tau_{Q}\sim (1/\tau_{t})(E_{w}/E_{dip})^{2}$. With $E_{w}\approx 0.03$~K and
$E_{dip}\approx 0.1$~K, it follows that $\tau_{Q}/\tau_{t}\approx 10$, in agreement with the experimental
magnetisation relaxation rate of $10^{-4}$~s$^{-1}$ observed at lowest $T$~\cite{Wernsdorfer00}. As argued by
Morello {\it et al.}~\cite{Morello04} and Baek {\it et al.}~\cite{Baek05}, the quantum tunneling fluctuations
of the cluster spins should set a $T$-independent lower limit to the longitudinal nuclear relaxation rate
$1/T_{1}^{n}$ equal to the quantum tunneling rate. Indeed, the calculated value of $1/\tau_{t}$ is precisely
that attained below about 0.5~K by the the experimental $1/T_{1}^{n}$ for $^{57}$Fe$_{8}$~\cite{Baek05}
(Fig.~3).

Our specific heat data thus neatly confirm the isotope effect predicted by the nuclear spin-mediated
tunneling (PS) model and previously seen in Ref.~\cite{Wernsdorfer00}, but in addition provide strong
evidence that in these incoherent quantum tunneling processes an efficient coupling to the phonons has to be
involved, i.e. relaxation of the cluster spins is actually to the lattice at a rate corresponding to the
square root relaxation, and not towards the nuclear spin-bath, as assumed in the PS model. However, the
problem is that, due to the strong average dipolar bias $E_{dip}\approx 0.1$~K, combined with the very small
$1/\tau_{t}\approx 1\times 10^{-3}~{\rm s}^{-1}$, a direct coupling of the tunneling levels to the phonons
via the usual spin-lattice interaction by phonon modulation of the crystal field leads to astronomically long
relaxation times (inset of Fig.~2). We thus propose a two-step relaxation process as an alternative. All data
show that in zero field, in order to flip the spin, we have to rely on the dynamic hyperfine interaction.
Through intercluster nuclear spin diffusion~\cite{Morello04} the hyperfine bias can fluctuate over an
appreciable part of the hyperfine split manifold. Since the width of this energy window $E_{w}$ is of the
same order (0.01 to 0.1~K) as the dipolar bias, near-resonant conditions for the tunneling levels can be met
by the combined action of both biases. Importantly, (i) when the spin flips, both the hyperfine field and the
dipolar bias acting on it change sign, implying that in this process energy can be interchanged between
nuclear spins and electron dipolar interaction reservoir, and (ii) the reversal of this dipolar bias is
instantaneous (picosec) compared to all relevant time scales and produces a temporary disturbance in the
local dipolar field distribution. But since the cluster spins are on the nodes of a crystal lattice, they are
coupled not only by dipolar but also by weak intercluster elastic (van der Waals) forces. Following general
arguments on energy and angular momentum conservation~\cite{Chudnovsky94}, this may result in a temporary
local lattice instability, similar as in a Franck-Condon type electronic transition associated with light
absorption by lattice defect or impurity states in solids~\cite{Maradudin66}. Thus an unbalance of the
interchanged hyperfine and dipolar energy quanta can be taken up by the lattice as potential energy in the
form of a local phonon mode, followed by dissipation into thermal phonons and simultaneous outward evolution
of the dipolar field distribution from the `defect'. The thermalisation of local vibrational modes due to
anharmonic processes has been studied by several authors~\cite{Maradudin66,Levinson80} and the associated
times are estimated to be of order $10^{3}$ to $10^{4}$~s at the low temperatures ($\approx 0.1$~K)
considered here~\cite{Levinson80}, faster indeed than the observed spin-lattice relaxation times (determined
by the tunneling rate).

The authors are indebted to J.F. Fern\'andez, A. Morello, S.I. Mukhin, P.C.E. Stamp, I.S. Tupitsyn and W.
Wernsdorfer for enlightening discussions. This work is part of the research program of the `Stichting FOM'
and is partially funded by the EC-RTN network ``QuEMolNa'' (No. MRTN-CT-2003-504880) and EC-Network of
Excellence ``MAGMANet'' (No. 515767-2). M.E. acknowledges MIUR for FIRB Project No. RBNE01YLKN.


\begin{thebibliography}{99}
\bibitem{lattice} The estimated $\theta_{D}\simeq 19$~K differs from $\theta_{D}\approx 34$~K
obtained in Ref.~\cite{Mettes01} from fits over a smaller $T$-range. From $\theta_{D}=19$~K we estimate the
sound velocity, $c_{s}=8\times 10^{2}$~m/s, which equals that found from the frequency-dependent
susceptibility ~\cite{Mettes01}.
\bibitem{note} To simulate the effect of intercluster dipolar coupling and hyperfine interactions,
we introduced a static $B_{z}=150$~G. For the model used, see \cite{Mettes01}.
\end{thebibliography}
\end{document}